\documentclass[letter,bibyear]{aa} % for the letters

\usepackage{natbib}
\usepackage{graphicx}
\usepackage{txfonts}
\newcommand{\jms}{J.~Mol.~Spectrosc.}   
  
\newcommand{\jmst}{J.~Mol.~Struct.}

\newcommand{\kms}{km s$^{-1}$}

\begin{document}

\title{Study of the  HCCNC and HNCCC isotopologs in TMC-1\thanks{Based on 
observations carried out
with the Yebes 40m telescope (projects 19A003,
20A014, 20D023, 21A011, 21D005 and 23A024). The 40m
radio telescope at Yebes Observatory is operated by the Spanish Geographic Institute
(IGN, Ministerio de Transportes, Movilidad y Agenda Urbana).}}

\author{
J.~Cernicharo\inst{1},
B.~Tercero\inst{2,3},
C.~Cabezas\inst{1},
M.~Ag\'undez\inst{1},
E.~Roueff\inst{4},
R.~Fuentetaja\inst{1},
N.~Marcelino\inst{2,3}, and
P.~de~Vicente\inst{3}
}

\institute{Dept. de Astrof\'isica Molecular, Instituto de F\'isica Fundamental (IFF-CSIC),
C/ Serrano 121, 28006 Madrid, Spain. \newline \email jose.cernicharo@csic.es, carlos.cabezas@csic.es
\and Observatorio Astron\'omico Nacional (OAN, IGN), C/ Alfonso XII, 3, 28014, Madrid, Spain.
\and Centro de Desarrollos Tecnol\'ogicos, Observatorio de Yebes (IGN), 19141 Yebes, Guadalajara, Spain.
\and LERMA, Observatoire de Paris, PSL Research University, CNRS, Sorbonne Universit\'e, F-92190 Meudon (France)
}

\date{Received; accepted}

\abstract{We present the detection of the three $^{13}$C isotopologs of HCCNC and HNCCC
toward TMC-1 using the QUIJOTE line survey. In addition, the D species has also been
detected for these two isomers of HCCCN, whereas the $^{15}$N isotopolog was only detected for HCCNC.
Using high-$J$ lines of HCCNC and HNCCC,
we were able to derive very precise rotational temperatures, column
densities, and subsequently the isotopic abundance ratios. We found that $^{12}$C/$^{13}$C is $\sim$90 for the
three possible substitutions in both isomers. These results are slightly different from what has been found for the
most abundant isomer HCCCN, for which abundances of 105, 95, and 66 were found
for each one of the three possible positions of $^{13}$C. 
The H/D abundance ratio was found to be 31$\pm$4 for HCCNC and of 53$\pm$6 for HNCCC. The latter
is similar to the H/D abundace ratio derived for HCCCN ($\sim$59).
The $^{14}$N/$^{15}$N isotopic abundance ratio in HCCNC is 243$\pm$24.
}

\titlerunning{Isotopologs of HCCNC and HNCCC}
\authorrunning{Cernicharo et al.}

\maketitle

\section{Introduction}
The ultra-sensitive line survey 
QUIJOTE\footnote{\textbf{Q}-band \textbf{U}ltrasensitive \textbf{I}nspection \textbf{J}ourney
to the \textbf{O}bscure \textbf{T}MC-1 \textbf{E}nvironment}
performed with the Yebes 40m radio telescope 
toward the prestellar cold core TMC-1
has enabled the unambiguous detection of near 50 molecules in the last three years
\citep[][and references therein]{Cernicharo2021,Cernicharo2023a}.
The sensitivity
of QUIJOTE is unprecedented, opening up new issues concerning the interpretation of the forest
of unknown lines present in sensitive spectral sweeps. In addition to the detection of new molecular species, which is the main goal of
the survey, we have to deal with the contribution of all isotopologs of any molecule producing line intensities
larger than 50 mK. Additionally, we must also  account for the presence of low-lying bending vibrational modes of abundant species 
\citep[see the case of the $\nu_{11}$ mode of C$_6$H,][]{Cernicharo2023a}. One of the advantages of
QUIJOTE is that it can provide a detection (for all isotopologs) of a molecule whose
emission is optically thin, thereby permitting a direct estimation of the isotopic abundances of species containing
C, O, S, N, and H.

The study of isotopic abundances as a function of the distance to the galactic center allows to trace 
stellar nucleosynthesis and constrain the chemical enrichment in our galaxy \citep[see, e.g.,][]{Yan2023}. 
Moreover, molecules are known to experience isotopic fractionation and this can be used to track the chemical 
evolution of molecular clouds and the transfer of chemical content to planetary systems \citep{Ceccarelli2014}. 
In this work, we report the detection and spectroscopic characterization of all isotopologs of HCCNC,
along with the first detection in space and first spectroscopic characterization, of the $^{13}$C isotopologs
of HNCCC. The results presented here will allow us to improve the chemical models dealing 
with isotopic fractionation in molecular clouds \citep{Roueff2015,Colzi2020,Loison2020,Sipila2023}. 
Although these models take into account $^{13}$C and $^{15}$N fractionation reactions with various 
degrees of approximation when including the dependence of the $^{13}$C position and adopted reactions, none 
of them deal with isotopic exchange reactions involving HNCCC and HCCNC.

\section{Observations}
The observational data used in this work are part of QUIJOTE \citep{Cernicharo2021}, 
a spectral line survey of TMC-1 in the Q-band carried out with the Yebes 40m telescope at 
the position $\alpha_{J2000}=4^{\rm h} 41^{\rm  m} 41.9^{\rm s}$ and $\delta_{J2000}=
+25^\circ 41' 27.0''$, corresponding to the cyanopolyyne peak (CP) in TMC-1. The receiver 
was built within the Nanocosmos project\footnote{\texttt{https://nanocosmos.iff.csic.es/}} 
and consists of two cold high-electron mobility transistor amplifiers covering the 
31.0-50.3 GHz band with horizontal and vertical polarizations. Receiver temperatures 
achieved in the 2019 and 2020 runs vary from 22 K at 32 GHz to 42 K at 50 GHz. Some 
power adaptation in the down-conversion chains have reduced the receiver temperatures 
during 2021 to 16\,K at 32 GHz and 30\,K at 50 GHz. The backends are 
$2\times8\times2.5$ GHz fast Fourier transform 
spectrometers with a spectral resolution of 38 kHz, providing the whole coverage 
of the Q-band in both polarizations.  A more detailed description of the system 
is given by \citet{Tercero2021}. 

\begin{figure}[h] 
\centering
\includegraphics[width=0.45\textwidth]{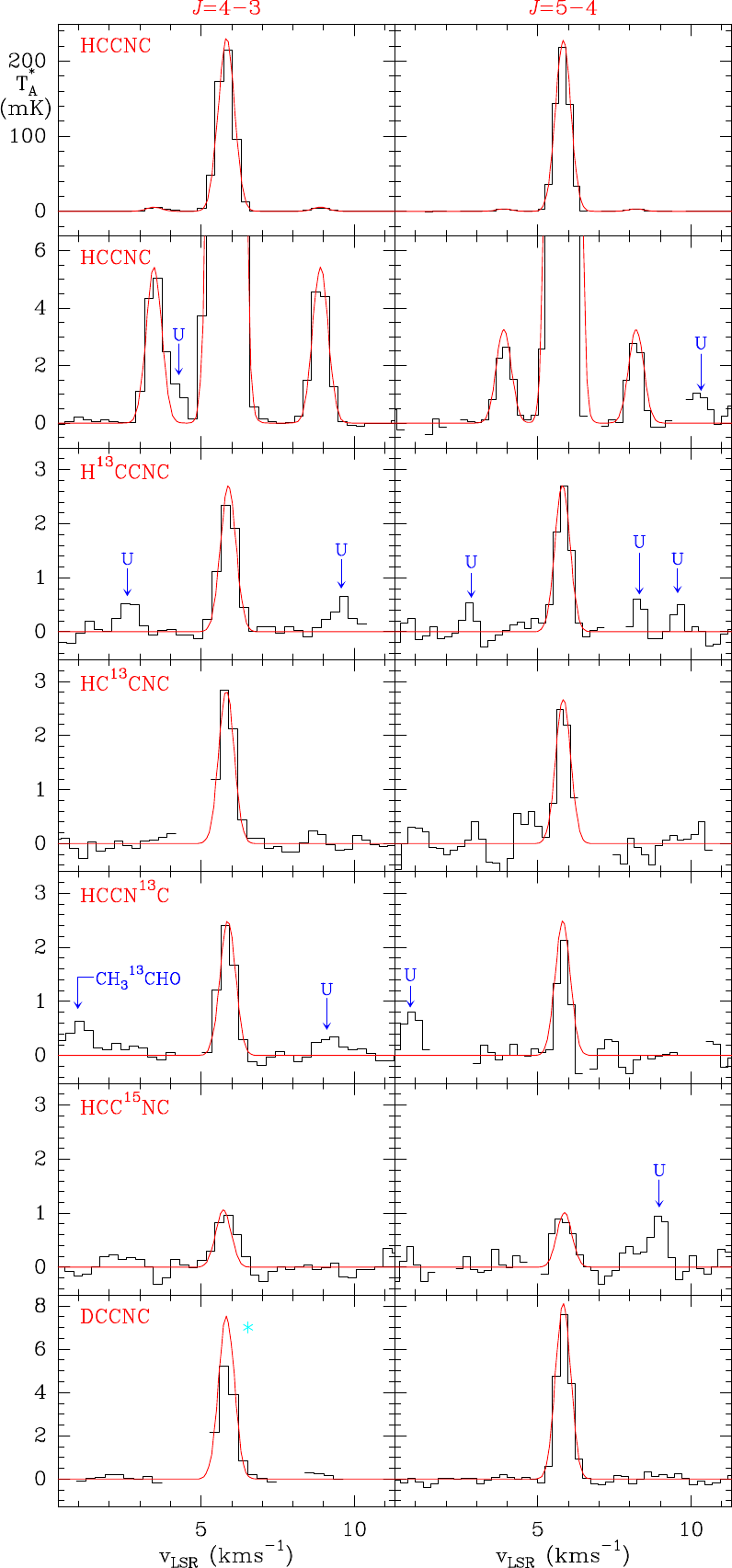}
\caption{
$J=4-3$ and $J=5-4$ transitions of the single substituted isotopologs of HCCNC.
The abscissa corresponds to the velocity with respect to the local standard of rest. The derived line parameters
are given in Table \ref{line_parameters}. The ordinate is the antenna temperature, corrected for 
atmospheric and telescope losses, in mK.
Blanked channels correspond to negative features 
produced when folding the frequency-switched data.
The red line shows the computed synthetic spectra for these lines (see Sect. \ref{results_trot}). The second 
row panels
correspond to a zoom in intensity of the lines of the main isotopolog (first row) to show the emission
from its weak hyperfine
components. The $J$=4-3 line
of DCCNC (marked with a cyan star) is affected by the negative features produced by the hyperfine structure of 
CH$_3$CN which is $\sim$8 MHz above in frequency. 
} 
\label{fig:HCCNC}
\end{figure}

The data of the QUIJOTE line survey presented here were gathered in several 
observing runs between November 2019 and July 2023.  
All observations are performed using frequency-switching observing mode with 
a frequency throw of 8 and 10 MHz. The total observing time on the source 
for data taken with frequency throws of 8 MHz and 10 MHz is 465 and 737 hours, 
respectively. Hence, the total observing time on source is 1202 hours. The 
measured sensitivity varies between 0.08 mK at 32 GHz and 0.2 mK at 49.5 GHz.  
The sensitivity of QUIJOTE is around 50 times better than that of previous 
line surveys in the Q band of TMC-1 \citep{Kaifu2004}. For each frequency 
throw, different local oscillator frequencies were used in order to remove 
possible side band effects in the down conversion chain. A detailed description 
of the QUIJOTE line survey is provided in \citet{Cernicharo2021}.
The data analysis procedure has been described by \citet{Cernicharo2022}.

The main beam efficiency measured during our observations in 2022 
varies from 0.66 at 32.4 GHz to 0.50 at 48.4 GHz \citep{Tercero2021} and can be given across the Q-Band by
$B_{\rm eff}$=0.797 exp[$-$($\nu$(GHz)/71.1)$^2$]. The
forward telescope efficiency is 0.97.
The telescope beam size at half-power intensity is 54.4$''$ at 32.4 GHz and 36.4$''$ 
at 48.4 GHz. 

Data for HCCNC and HNCCC in the millimeter domain have been taken with the IRAM 30m 
telescope and consist of a 3mm line survey 
that covers the full available band at the telescope, between 71.6 GHz and 
117.6 GHz. The EMIR E0 receiver was connected to the Fourier transform spectrometers 
(FTS) in its narrow mode, which provide a spectral resolution of 49 kHz and a total 
bandwidth of 7.2 GHz per spectral setup. Observations were performed in several runs. 
Between January and May 2012, we completed the scan 82.5--117.6 GHz \citep{Cernicharo2012b}. 
In August 2018, after the upgrade of the E090 receiver, we extended the survey down 
to 71.6 GHz. More recent high sensitivity observations in 2021 have been used to improve 
the signal to noise ratio (S/N) in several frequency windows \citep{Agundez2022,Cabezas2022}. 
The final 3mm line survey has a sensitivity of 2-10 mK. However, at some selected 
frequencies, the sensitivity is as low as 0.6 mK. All the observations were performed using the 
frequency-switching method with a frequency throw of 7.14 MHz. The IRAM 30m beam 
varies between 34$''$ and 21$''$ at 72 GHz and 117 GHz, respectively, while the beam 
efficiency takes values of 0.83 and 0.78 at the same frequencies, following the 
relation: B$_{eff}$= 0.871 exp[$-$($\nu$(GHz)/359)$^2$]. The forward efficiency at 3mm is 0.95.

The intensity scale 
utilized in this study is the antenna temperature ($T_A^*$). 
The calibration was performed using two absorbers at 
different temperatures and the atmospheric transmission model ATM \citep{Cernicharo1985, 
Pardo2001}. The absolute calibration uncertainty is 10$\%$. However, the relative
calibration between lines within the QUIJOTE survey is certainly better as all them
are observed simultaneously.
The data were analyzed with the GILDAS package\footnote{\texttt{http://www.iram.fr/IRAMFR/GILDAS}}.

\vspace{-0.3cm}
\section{Results}\label{sec:results}

Line identification in this work has been performed using the MADEX code \citep{Cernicharo2012}, and the CDMS and JPL catalogues \citep{Muller2005,Pickett1998}.
The references and laboratory data used by MADEX in its spectroscopic predictions are described below.

\subsection{Spectroscopy of HCCNC}\label{spec_HCCNC}
HCCNC was detected toward TMC-1 by \citet{Kawaguchi1992a}.
The dipole moment is 2.93\,D \citep{Kruger1991}, which is lower than that of
HCCCN.
\citet{Guarnieri1992} observed
rotational transitions of this species up to $J_u$=33. Hence, the rotational and distortion
constants have been well established. However, the hyperfine structure was observed only for the
$J$=1-0 line \citep{Kruger1992}. The hyperfine structure of the 
$J$=4-3 and $J$=5-4 lines was observed toward TMC-1 with a previous version of the QUIJOTE line survey 
\citep{Cernicharo2020}. 
We remeasured the frequencies of these lines with the present QUIJOTE data (see Table \ref{line_parameters}). 
In addition, the $J$=8-7 up to $J$=11-10 lines of HCCNC were observed with the IRAM 30m radio telescope.
They are shown in Fig. \ref{fig:3mm} and their line parameters are given in Table \ref{line_parameters}. 
A fit to all the laboratory and space data was performed using
the standard Hamiltonian of a linear molecule with a hyperfine structure. The results
are given in Table \ref{rotationalconstants}. 
Therefore, the predictions for HCCNC, including the hyperfine structure, are reliable up to very high-$J$.
In addition, DCCNC was also observed in the laboratory up to $J_u$=51 \citep{Huckauf1998}. The
situation was completely different for the $^{13}$C and $^{15}$N isotopologs, for which
only the $J$=1-0 transition has been observed in the laboratory \citep{Kruger1992}. 
All these isotopologs
were observed and characterized spectroscopically by \citet{Cernicharo2020}, who observed
their $J$=4-3 and $J$=5-4 transitions toward TMC-1. The present sensitivity of QUIJOTE is
much better than that of 2020 and the frequency of the lines has been measured again in
order to improve the rotational constants of these species. The observed lines of HCCNC,
H$^{13}$CCNC, HC$^{13}$CNC, HCCN$^{13}$C, HCC$^{15}$NC, and DCCNC are shown in Fig.
\ref{fig:HCCNC} and their line parameters are given in Table \ref{line_parameters}.

\begin{figure}[h] 
\centering
\includegraphics[width=0.485\textwidth]{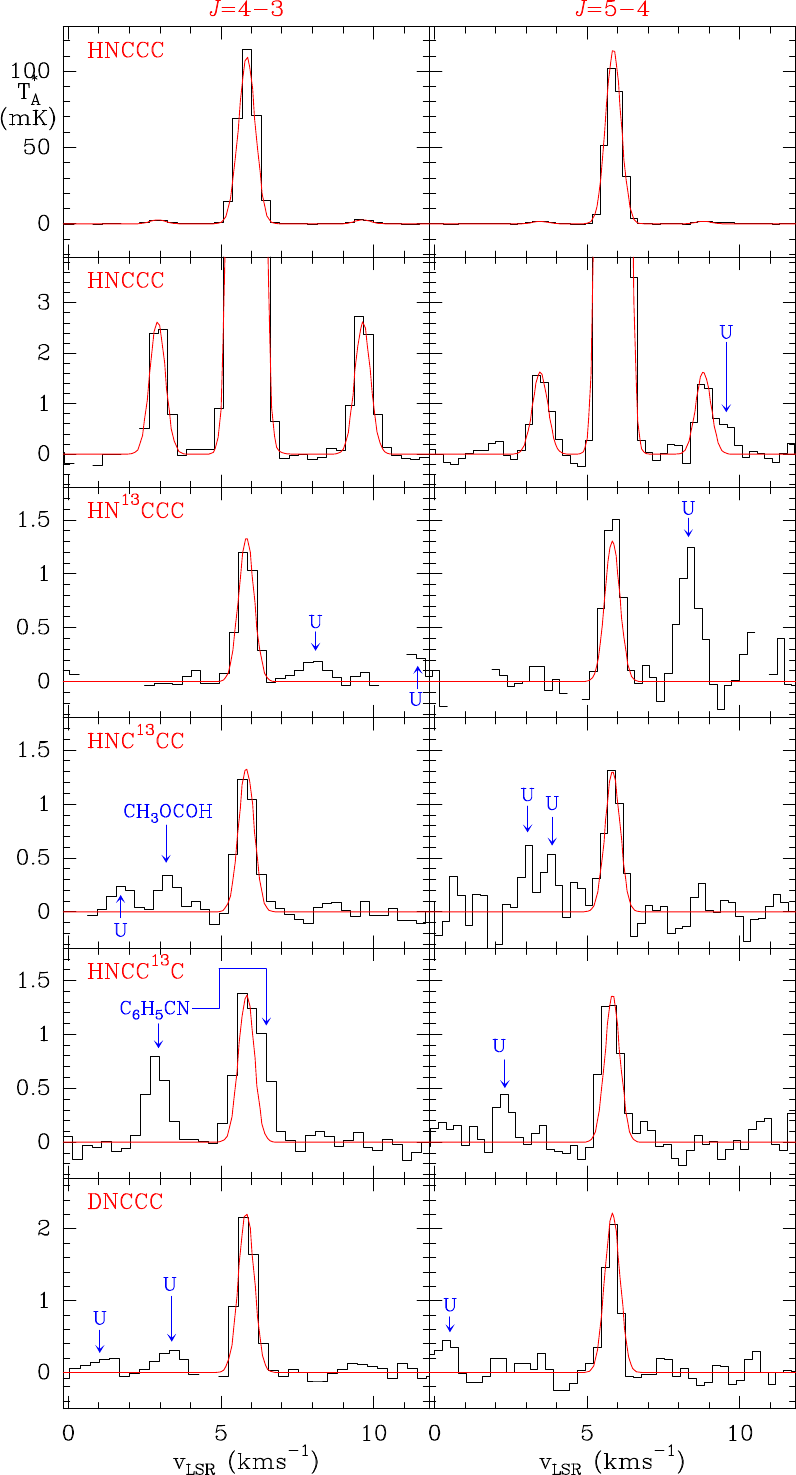}
\caption{Same as Fig. \ref{fig:HCCNC}, but for HNCCC.
The red line shows the computed synthetic spectra for these lines (see Sect. \ref{results_trot}). The second row
correspond to a zoom in intensity of the lines of the main isotopolog (first row).
The derived line parameters
are given in Table \ref{line_parameters}.
} 
\label{fig:HNCCC}
\end{figure}
\begin{table*}
%\small
\centering
\caption{Derived rotational constants and column densities for the isotopologs of HCCNC and HNCCC}
\label{rotationalconstants}
\centering
%\begin{tabular}{{|lccccc|c|c|}}
\begin{tabular}{{lccccccc}}
\hline
Species             & $B$ (MHz)     & $D$ (kHz)  &  $eQq$ (MHz) &   $J_{max}$ &$\sigma$ (kHz) & N (cm$^{-2}$) & Notes\\
\hline                               
HCCNC               & 4967.83827(12)  & 0.62694(14)& 0.94628(81)&33 &20.0 & (3.4$\pm$0.10)$\times$10$^{12}$ &A\\%V
H$^{13}$CCNC        & 4813.11872(66)  & 0.559(41)  & 0.9429(39) & 5 & 6.2 & (3.7$\pm$0.15)$\times$10$^{10}$ &B\\
HC$^{13}$CNC        & 4949.44614(22)  & 0.605(14)  & 0.9457(13) & 5 & 0.7 & (3.7$\pm$0.15)$\times$10$^{10}$ &B\\
HCCN$^{13}$C        & 4804.87363(22)  & 0.555(13)  & 0.9447(12) & 5 & 4.1 & (3.4$\pm$0.15)$\times$10$^{10}$ &B\\
HCC$^{15}$NC        & 4944.73993(78)  & 0.516(33)  &            & 5 &14.9 & (1.4$\pm$0.15)$\times$10$^{10}$ &B\\
DCCNC               & 4598.28896(03)  & 0.518(0)   &            &51 & 3.3 & (1.1$\pm$0.15)$\times$10$^{11}$ &C\\
\hline
HNCCC               & 4668.33752(25)  & 0.62165(72)& 1.0995(48) &31 &19.9 & (5.8$\pm$0.15)$\times$10$^{11}$ &D\\
HN$^{13}$CCC        & 4655.7452(50)   & 0.61       &            & 5 & 1.0 & (6.2$\pm$0.20)$\times$10$^{09}$ &E\\
HNC$^{13}$CC        & 4643.5839(50)   & 0.61       &            & 5 & 1.0 & (6.2$\pm$0.20)$\times$10$^{09}$ &E\\
HNCC$^{13}$C        & 4500.9332(50)   & 0.57       &            & 5 & 14.0& (6.6$\pm$0.20)$\times$10$^{09}$ &E\\
DNCCC               & 4400.5916(50)   & 0.52       & 1.169(19)  & 5 & 22.3& (1.1$\pm$0.20)$\times$10$^{10}$ &F\\
\hline
\end{tabular}
\tablefoot{
\tablefoottext{A}{Fit to the laboratory lines measured by
\citet{Guarnieri1992} and \citet{Kruger1992}, and to the 
hyperfine components of the $J$=4-3 and 5-4 lines observed with QUIJOTE (see Table \ref{line_parameters}).}
\tablefoottext{B}{Fit to the laboratory $J$=1-0 hyperfine components measured by
\citet{Kruger1992} and to the main
component of the $J$=4-3 and 5-4 lines observed with QUIJOTE (see Table \ref{line_parameters}).}
\tablefoottext{C}{Fit to the laboratory lines measured by
\citet{Huckauf1998}.}
\tablefoottext{D}{Fit to the laboratory lines measured by 
\citet{Kawaguchi1992b}, \citet{Hirahara1993}, and \citet{Vastel2018}.
The hyperfine components of the $J$=4-3 and 5-4 lines measured with QUIJOTE
have been included in the fit. We have also added the frequencies of the
$J$=8-7, 9-8 and 10-9 lines measured with the IRAM 30m radio telescope (see Table \ref{line_parameters}).
The distortion constant $H$ has been derived to be 
(1.57$\pm$0.48)$\times$10$^{-9}$ MHz. Its inclusion in the fit reduces the
standard deviation from 23.3 kHz to 19.9 kHz.}
\tablefoottext{E}{Fit to the two lines measured toward TMC-1. The distortion
constant has been fixed for each isotopolog (see Sect. \ref{spec_HNCCC}) and the uncertainty on $B$ has been assumed to be
5 kHz.}
\tablefoottext{F}{Fit to the $J$=1-0 and 2-1 lines measured by \citet{Hirahara1993}
and to the $J$=4-3 and 5-4 lines observed toward TMC-1 in this work.}
}
\end{table*}

\begin{table}
\small
\centering
\caption{Isotopic abundance ratios}
\label{iso_ratios}
\centering
\begin{tabular}{{|lccccc|c|c|}}
\hline
Species  & $^{12}$C/$^{13}$C$_1$  & $^{12}$C/$^{13}$C$_2$  &  $^{12}$C/$^{13}$C$_3$ &   H/D    & N/$^{15}$N \\
\hline
HCCNC    &      93$\pm$6          &  93$\pm$6              &  100$\pm$7           &   31$\pm$5 &  243$\pm$24\\
HNCCC    &      94$\pm$5          &  94$\pm$5              &   88$\pm$5           &   53$\pm$6 &            \\
HCCCN$^A$&     106$\pm$4          &  95$\pm$2              &   69$\pm$1           &   57$\pm$2 &  317$\pm$4 \\
\hline
\end{tabular}
\tablefoot{
\tablefoottext{A}{From \citet{Tercero2024}.}
}
\end{table}
\normalsize

\subsection{Spectroscopy of HNCCC}\label{spec_HNCCC}
HNCCC was discovered in TMC-1 by \citet{Kawaguchi1992b} through the observation of  three unknown lines
in the Nobeyama spectral  line survey of TMC-1 \citep{Kaifu2004} that they assigned to this isomer of HCCCN. 
The dipole moment of this molecule has been computed through ab initio
calculations to be 5.67\,D \citep{Botschwina1992}. The molecule was observed in the laboratory by \citet{Hirahara1993}
confirming the assignment of \citet{Kawaguchi1992b}. These observations included the hyperfine components of the $J$=1-0 and $J$=2-1 transitions.
Laboratory line frequencies up to $J$=31-30 have been reported by \citet{Vastel2018}. Several lines at 3mm have been also
observed with our data gathered  with the IRAM 30m radio telescope (see Fig. \ref{fig:3mm}). In addition, we have observed the hyperfine structure of the $J$=4-3
and $J$=5-4 lines with the QUIJOTE line survey (see Fig. \ref{fig:HNCCC}). The derived line parameters for all observed transitions of HNCCC 
are given in Table \ref{line_parameters}.
All the spectroscopic data from the laboratory and our space observations were fitted with the same Hamiltonian than for HCCNC. 
The results are given in Table \ref{rotationalconstants}. Unlike the case of HCCNC, the only spectral information for the isotopologs
of HNCCC concerns DNCCC \citep{Hirahara1993}. This isotopolog was observed in TMC-1 by \citet{Cernicharo2020}. Improved frequencies for DNCCC are
reported in Table \ref{line_parameters} and its derived rotational constants are given in Table \ref{rotationalconstants}. 
Assuming that the emission of HNCCC is not affected by line opacity, then we could expect to detect the lines from the $^{13}$C isotopologs 
at the level of 1-1.5 mK. For H$^{15}$NCCC, the same lines should be at the level of 0.5 mK  (assuming the HCCNC/HCC$^{15}$NC abundance ratio). 
The expected molecular constants, $B_0$ and $D_0$, of the 
isotopologs have been computed through ab initio calculations using 
B$_{0,exp}$(HNCCC)/B$_{e,cal}$(HNCCC) as the scaling factor.
%have been computed through ab initio calculations using HNCCC as reference. 
We estimated $B_0$$\sim$ 4655.5, 4644.0, 4501.0, and 4545.4 MHz for
HN$^{13}$CCC, HNC$^{13}$CC, HNCC$^{13}$C, and H$^{15}$NCCC, respectively. 
Three pairs of lines with
similar intensities, as is the case for the isotopologs of HCCNC, have been easily found with a near perfect 
harmonic frequency relation 5/4 in our data close in frequency to our predictions.
The deviation from this relation is 3$\times$10$^{-6}$ and fits the expected contribution of the distortion constant ($\sim$0.6 kHz).
They correspond to the three $^{13}$C isotopologs and they are shown in Fig. \ref{fig:HNCCC}. Their line parameters
are given in Table \ref{line_parameters}. As only two lines are available for each isotopolog, we have adopted
a distortion constant of 0.61 kHz for HN$^{13}$CCC and HNC$^{13}$CC, and of 0.57 kHz for HNCC$^{13}$C. The derived rotational constants, for which
we estimated a conservative uncertainty of 5 kHz,  are given in Table \ref{rotationalconstants}. We have not been able to find a similar pair of lines
that could be assigned to the $^{15}$N isotopolog because too many features do appear at the expected intensities of 0.5 mK.
Thus, we report the first detection in space of the $^{13}$C isotopologs of HNCCC and the determination of
their rotational constant.

\subsection{Rotational temperatures and column densities}\label{results_trot}
We analyzed the intensities of the lines of HCCNC and HNCCC using a rotational diagram to derive
the rotational temperatures of these molecules. We assumed a source of uniform brightness temperature and a radius of
40$''$ \citep[][Fuentetaja et al. 2023, in prep]{Fosse2001} to estimate the dilution factor for each transition. 
The line width adopted in the models to compute the synthetic spectra is 0.6\,km\,s$^{-1}$ for all lines.
The results of the rotational diagram for HCCNC are T$_{rot}$=5.5$\pm$0.1\,K and $N$=(3.4$\pm$0.1)$\times$10$^{12}$ cm$^{-2}$, while for HNCCC, we obtained T$_{rot}$=4.5$\pm$0.2\,K and $N$=(5.8$\pm$0.2)$\times$10$^{11}$ cm$^{-2}$.
The corresponding synthetic spectra are shown in Fig. \ref{fig:HCCNC} for HCCNC and in Fig. \ref{fig:HNCCC}
for HNCCC. We note that the weak hyperfine satellite lines of HCCNC and HNCCC are very well
reproduced with the derived values for the column densities. 
These hyperfine components represent 2\% 
and 1.3\% of the total intensity of the  $J$=4-3 and 5-4 transitions, respectively. The 
fact that the modeled intensities for the weak and strong hyperfine components are in perfect agreement with
the observations indicates that the emission of the $J$=4-3 and 5-4 transitions of HCCNC. In addition, HNCCC is mostly optically thin. Hence, the derived column densities can be used to obtain isotopic abundance ratios.
By adopting the rotational temperatures derived for HCCNC and HNCCC for all their isotopologs, we obtained the column densities given in Table \ref{rotationalconstants}. The match between computed and 
observed spectra is excellent (see Figures \ref{fig:HCCNC} and \ref{fig:HNCCC}) and the derived isotopic abundance ratios are given in Table \ref{iso_ratios}.
We note that the synthetic spectra derived from the assumption of a constant rotational temperature
reproduce the observed spectra at 3mm for both species rather well. Nevertheless, the column density of
HNCCC has to be reduced by a factor 1.5 to reproduce the observed intensities of this species at 3mm.

Collisional rates for HCCNC and HNCCC with $p$-H$_2$ 
and $o$-H$_2$ have been calculated by \citet{Bop2021}. We used the large velocity gradient approximation 
(LVG), described by \citet{Goldreich1974} and implemented
in the MADEX code \citep{Cernicharo2012}, to model the line profiles of the $J$=4-3 to $J$=11-10 lines of 
both isomers. We explored the density and the column density that produce the best fit to the $J$=8-7 to
$J$=11-10 lines of these species using $p$-H$_2$ as collider. For HCCNC the best fit is obtained for 
n(H$_2$)=(7.0$\pm$0.3)$\times$10$^{3}$ cm$^{-3}$ and $N$=(2.7$\pm$0.2)$\times$10$^{12}$ cm$^{-2}$. 
The excitation temperatures vary between 7.3\,K for the $J$=4-3 transition to 4.7\,K for the $J$=11-10 one.
The generated
synthetic spectra are shown in blue in Fig. \ref{fig:3mm}. These synthetic spectra agree rather well with
those obtained  for a constant rotational temperature of 5.5\,K and $N$=3.4$\times$10$^{12}$ cm$^{-2}$ 
(synthetic spectra shown in red in Fig. \ref{fig:3mm}).
However, this fit to the volume density and column density to the 3mm lines does not adequately reproduce the 
observations of the $J$=4-3 and $J$=5-4 lines. The differences between the two methods are probably related
to the density structure of the source and to the accuracy of the collisional rates. We note that the
derived density is lower than that derived from other molecules in TMC-1. We checked the effect of
using $o$-H$_2$ as collider and found similar results.
In the case of HNCCC, our LVG calculations indicate that the intensity of the 3mm lines are well reproduced,
with n(H$_2$)=1.5$\times$10$^4$ cm$^{-3}$ and a column density of 3.8$\times$10$^{11}$. 
The excitation temperatures vary between 5.6\,K for the $J$=4-3 transition to 4.0\,K for the $J$=11-10 one.
The computed synthetic
spectra are shown in blue in Fig. \ref{fig:3mm} and agree well with those obtained from a constant rotational
temperature (red curve in Fig. \ref{fig:3mm}). It is worth  noting that in this case the derived 
volume density is in better agreement with the
values derived from other species. We note also that as for HCCNC, the best column density obtained from the 
LVG analysis is lower than that
obtained by the rotational diagram. The most plausible explanation for these differences in both
isomers is related to
the cloud density structure, that is, the dense core of TMC-1 can not be treated as an object of uniform density.
Moreover, it is also likely that molecular abundances vary across the core.

\section{Discussion}\label{sec:discussion}
For the following discussion, we adopt the values derived above for the column densities using an
uniform rotational temperature for all observed transitions. 
Taking the column density of HCCCN derived by Tercero et al. (2023, in prep.) of 1.9$\times$10$^{14}$ cm$^{-2}$, the relative abundance ratios are 56, 328, and 6 for HCCCN/HCCNC, HCCCN/HNCCC, and
HCCNC/HNCCC, respectively. These values agree, within a factor of two, with those obtained from less sensitive line surveys of TMC-1 \citep{Gratier2016}. The derived abundance for the isotopologs of
HCCNC, HNCCC, and HCCCN permits us to study the isotopic $^{12}$C/$^{13}$C abundance ratio for the three possible
substitutions. The results are given in Table \ref{iso_ratios}. We found that for HCCNC and HNCCC this 
isotopic ratio does not depend on the substituted carbon, being very similar to the solar one and  
slightly larger than that of the local ISM of 59-76 \citep{Lucas1998,Wilson1999,Milam2005,Sheffer2007,Stahl2008,Ritchey2011}.
The results from Tercero et al. (2023, in prep) for HCCCN also indicate that for the most abundant isomer, the
only isotopolog whose ratio is slightly different is HCC$^{13}$CN. Table \ref{iso_ratios} presents a summary of
these ratios.

The $^{13}$C isotopic ratios derived for HCCCN in TMC-1 are in agreement with previous determinations in other cold 
dense clouds, such as L1521B, L134N \citep{Taniguchi2017}, L1527 \citep{Yoshida2019}, and L483 \citep{Agundez2019}; more specifically, HCC$^{13}$CN is somewhat more abundant than the other two $^{13}$C isotopologs. The $^{12}$C/$^{13}$C 
ratio for HCC$^{13}$CN in TMC-1 is consistent with the local ISM $^{12}$C/$^{13}$C ratio, while the isotopologs 
H$^{13}$CCCN and HC$^{13}$CCN are somewhat diluted in $^{13}$C. This behavior is reproduced by the chemical model 
of \cite{Loison2020}, assuming that C$_3$ reacts with oxygen atoms, although we caution that there are 
uncertainties in some of the relevant chemical reactions. In the cases of HCCNC and HNCCC, the three $^{13}$C 
isotopologs show similar abundances, somewhat above the local ISM $^{12}$C/$^{13}$C ratio. A slight dilution 
in $^{13}$C was therefore found for the three isotopologs of HCCNC and HNCCC. 
%It would be interesting to compare 
%these results with predictions from chemical models \citep{Roueff2015,Loison2020}, which for the moment do not 
%consider these molecules.

The dependence of the isotopic ratio between HC$_3$N and carbon substituted HC$_3$N on the position 
of the $^{13}$C 
was previously discussed by \citet{Taniguchi2016,Taniguchi2017}. They concluded that the main 
formation of HC$_3$N comes from the neutral-neutral reaction between acetylene C$_2$H$_2$ and CN.
The $^{13}$C substituted HC$_3$N has been considered in the fractionation models of \citet{Loison2020}, who obtained a satisfactory 
agreement with observations when introducing a moderate value for the C$_3$ and O reaction rate coefficient. The most recent paper by \citet{Sipila2023}  
only introduces $^{13}$C fractionation reactions for CCH, C$_3$, and c-C$_3$H$_2$. The three $^{13}$C isotopomers of HC$_3$N 
thus have  equal abundances over the full time dependence in the corresponding models.

The carbon and nitrogen isotopic chemistry of HCCNC and HNCCC has not seen very much interest until now and no chemical model 
has been performed to predict their isotopic ratios as yet. The deuterated species were detected by \citet{Cernicharo2020}
and the abundance ratios were reported and modeled in 
\citet{Cabezas2021} with reasonable success. Interestingly, the recent detection of HCCNCH$^+$ by \citet{Agundez2022}, showing a high HCCNCH$^+$ over HCCNC ratio (on the order of 10$^{-2}$), is surprisingly well reproduced in the associated 
chemical model. This is contrary to the results seen for most other protonated molecules. Following the KIDA database suggestion \citep{Wakelam2012}, 
we find that HCCNCH$^+$ is efficiently formed through the C$^+$ + CH$_3$CN reaction that can return HCCNC through dissociative recombination 
amongst other products. This mechanism prevails over the other proton transfer reactions in our chemical model. Such an ion-molecule 
scheme may lead to different isotopic ratios than the  neutral-neutral formation channels. 
Further studies on these topics are  desirable to build on these results.

\begin{acknowledgements}

We thank Ministerio de Ciencia e Innovaci\'on of Spain (MICIU) for funding support through projects
PID2019-106110GB-I00, PID2019-107115GB-C21 / AEI / 10.13039/501100011033, and
PID2019-106235GB-I00. We also thank ERC for funding
through grant ERC-2013-Syg-610256-NANOCOSMOS.

\end{acknowledgements}

%\normalsize
%\onecolumn
\begin{appendix}
\section{Line parameters}\label{app:lineparameters}
\begin{figure}[h] 
\centering
\includegraphics[width=0.485\textwidth]{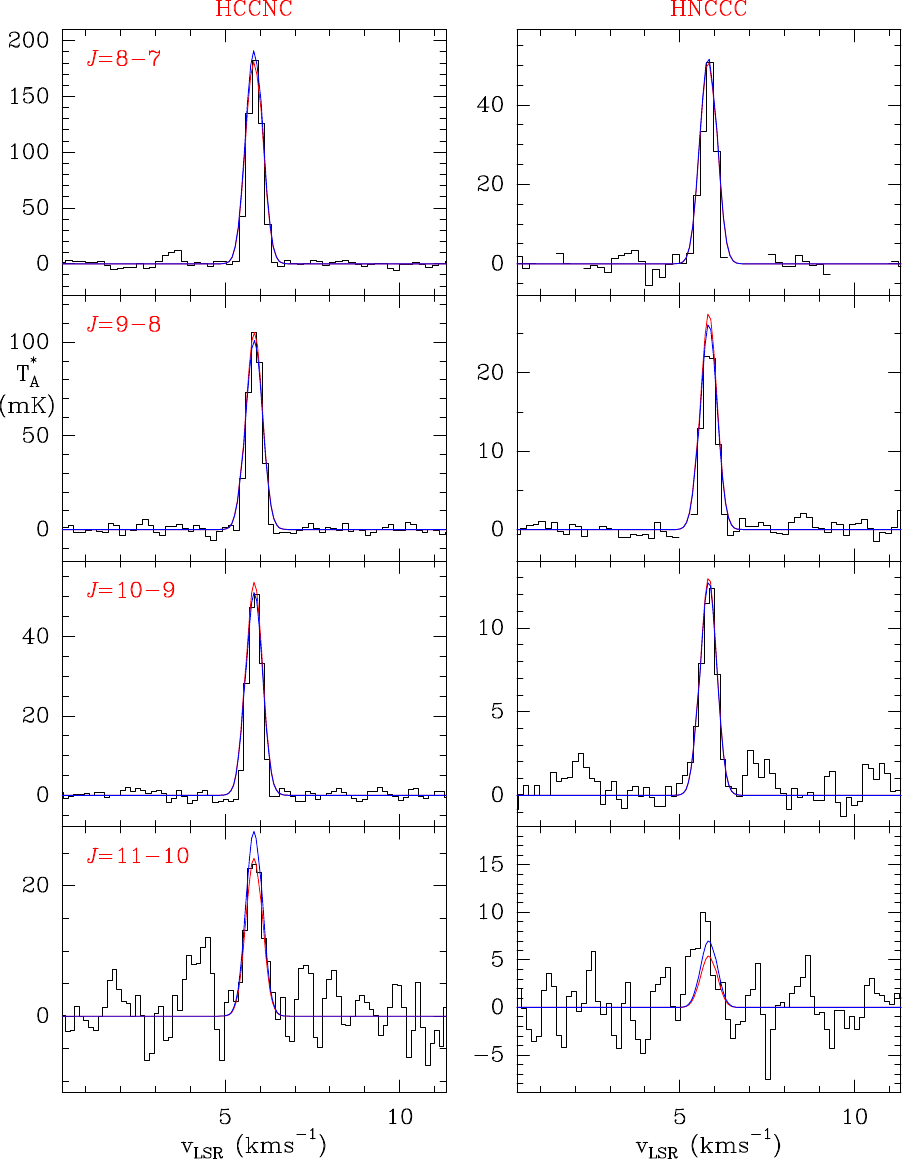}
\caption{Observed lines in the 3mm domain of HCCNC (left panels) and HNCCC (right panels).
The abscissa and ordinate are those of Fig. \ref{fig:HCCNC}.
The computed synthetic spectra adopting a constant rotational temperature are shown in red, whereas those obtained
from the LVG model are shown in blue (see Sect. \ref{results_trot}).
The derived line parameters
are given in Table \ref{line_parameters}.
} 
\label{fig:3mm}
\end{figure}

Line parameters for all observed transitions were derived by fitting a Gaussian line profile to them
using the GILDAS package. A
velocity range of $\pm$20\,\kms\, around each feature was considered for the fit after a polynomial 
baseline was removed. Negative features produced in the folding of the frequency switching data were blanked
before baseline removal. The derived line parameters for all observed lines are
given in Table \ref{line_parameters}. The $J$=4-3 and $J$=5-4 lines of HCCNC are shown in Fig. \ref{fig:HCCNC}
and those of HNCCC in Fig. \ref{fig:HNCCC}. The $J$=8-7 to $J$=11-10 lines of HCCNC and HNCCC are
shown in Fig. \ref{fig:3mm}.

\begin{table*}[h]
\centering
\caption{Observed line parameters for the isotopologs of HCCNC and HNCCC}
\label{line_parameters}
\begin{tabular}{lcccrccr}
\hline
Isotopolog    & $J'-J$''$$          & $F'-F$''$$& $\nu_{rest}$~$^a$ & $\int T_A^* dv$~$^b$ & v$_{LSR}$       & $\Delta v$~$^c$ & $T_A^*$~$^d$ \\
                &                      & & (MHz)              & (mK\,km\,s$^{-1}$)  & (km\,s$^{-1}$)  & (km\,s$^{-1}$)  & (mK) \\
\hline
HCCNC       & 4-3& 3-3 &39742.141$\pm$0.010&  3.46$\pm$0.06& 5.83         & 0.63$\pm$0.01&  5.20$\pm$0.08\\ %3E12
            &    &Main &39742.551$\pm$0.010&158.68$\pm$0.06& 5.83         & 0.66$\pm$0.01&225.26$\pm$0.08\\
            &    &4-4  &39742.854$\pm$0.010&  4.17$\pm$0.07& 5.83         & 0.76$\pm$0.02&  5.15$\pm$0.08\\
            & 5-4& 4-4 &49677.692$\pm$0.010&  1.70$\pm$0.18& 5.83         & 0.52$\pm$0.06&  3.05$\pm$0.35\\
            &    &Main &49678.073$\pm$0.010&133.98$\pm$0.16& 5.83         & 0.57$\pm$0.01&220.85$\pm$0.35\\
            &    &5-5  &49678.380$\pm$0.010&  1.79$\pm$0.19& 5.83         & 0.61$\pm$0.07&  2.76$\pm$0.35\\
            & 8-7&     &79484.128$\pm$0.002& 98.19$\pm$1.20& 5.81$\pm$0.01& 0.50$\pm$0.01&186.30$\pm$2.12\\
            & 9-8&     &89419.261$\pm$0.002& 55.74$\pm$0.88& 5.82$\pm$0.01& 0.48$\pm$0.01&109.13$\pm$2.10\\
            &10-9&     &99354.258$\pm$0.002& 26.27$\pm$0.40& 5.82$\pm$0.01& 0.46$\pm$0.01& 53.66$\pm$0.99\\
            &11-10&   &109289.104$\pm$0.002& 15.40$\pm$1.58& 5.81$\pm$0.03& 0.59$\pm$0.07& 24.25$\pm$3.46\\
H$^{13}$CCNC& 4-3&     &38504.814$\pm$0.010&  1.82$\pm$0.14& 5.83         & 0.69$\pm$0.06&  2.47$\pm$0.10\\ %3.7E10 R=81.1
            & 5-4&     &48130.904$\pm$0.010&  1.63$\pm$0.12& 5.83         & 0.56$\pm$0.05&  2.73$\pm$0.17\\
HC$^{13}$CNC& 4-3&     &39595.414$\pm$0.010&  1.97$\pm$0.08& 5.83         & 0.63$\pm$0.03&  2.95$\pm$0.11\\ %3.7E10 R=81.1
            & 5-4&     &49494.159$\pm$0.010&  1.50$\pm$0.17& 5.83         & 0.53$\pm$0.07&  2.67$\pm$0.25\\
HCCN$^{13}$C& 4-3&     &38438.852$\pm$0.010&  1.72$\pm$0.10& 5.83         & 0.65$\pm$0.04&  2.49$\pm$0.10\\ %3.4E10 R=88.2
            & 5-4&     &48048.456$\pm$0.010&  1.08$\pm$0.11& 5.83         & 0.46$\pm$0.05&  2.21$\pm$0.17\\
HCC$^{15}$NC& 4-3&     &39557.774$\pm$0.010&  0.85$\pm$0.10& 5.83         & 0.81$\pm$0.10&  9.94$\pm$0.13\\ %1.4E10 R=214.3
            & 5-4&     &49447.148$\pm$0.010&  0.73$\pm$0.11& 5.83         & 0.70$\pm$0.11&  9.85$\pm$0.20\\
DCCNC       & 4-3&     &36786.179$\pm$0.001&  3.93$\pm$0.12& 5.83$\pm$0.01& 0.68$\pm$0.02&  5.40$\pm$0.16\\ %1.1E11 R=27.3
            & 5-4&     &45982.631$\pm$0.001&  4.62$\pm$0.10& 5.84$\pm$0.01& 0.56$\pm$0.01&  7.71$\pm$0.15\\
\\
\hline
\\
HNCCC       & 4-3& 3-3 &37346.067$\pm$0.010&  2.16$\pm$0.07& 5.83         & 0.69$\pm$0.02&  2.94$\pm$0.09\\ %5.8E11
            &    & Main&37346.541$\pm$0.010& 88.04$\pm$0.07& 5.83         & 0.72$\pm$0.01&114.52$\pm$0.09\\
            &    & 4-4 &37346.902$\pm$0.010&  1.91$\pm$0.02& 5.83         & 0.62$\pm$0.02&  2.88$\pm$0.09\\
            & 5-4& 4-4 &46682.600$\pm$0.010&  1.39$\pm$0.13& 5.83         & 0.98$\pm$0.12&  1.33$\pm$0.15\\
            &    & Main&46683.063$\pm$0.010& 69.54$\pm$0.14& 5.83         & 0.61$\pm$0.01&107.31$\pm$0.15\\
            &    & 5-5 &46683.435$\pm$0.010&  1.17$\pm$0.15& 5.83         & 0.68$\pm$0.10&  1.63$\pm$0.15\\
            & 8-7&     &74692.130$\pm$0.010& 28.86$\pm$2.19& 5.83         & 0.52$\pm$0.05& 51.93$\pm$4.08\\ 
            & 9-8&     &84028.265$\pm$0.010& 12.71$\pm$0.37& 5.83         & 0.49$\pm$0.02& 24.22$\pm$0.75\\
            &10-9&     &93364.268$\pm$0.010&  7.53$\pm$0.38& 5.83         & 0.57$\pm$0.03& 12.40$\pm$0.74\\
            &11-10&   &102700.117$\pm$0.001&  6.43$\pm$1.39& 6.03$\pm$0.07& 0.67$\pm$0.16&  9.10$\pm$3.06\\ 
HN$^{13}$CCC& 4-3&     &37245.805$\pm$0.010&  0.87$\pm$0.08& 5.83         & 0.65$\pm$0.07&  1.26$\pm$0.09\\ %6.2E9 R=93.5
            & 5-4&     &46557.148$\pm$0.010&  1.04$\pm$0.14& 5.83         & 0.71$\pm$0.09&  1.69$\pm$0.22$^e$\\
HNC$^{13}$CC& 4-3&     &37148.515$\pm$0.010&  1.00$\pm$0.07& 5.83         & 0.72$\pm$0.06&  1.31$\pm$0.07\\ %6.2E9 R=93.5
            & 5-4&     &46435.534$\pm$0.010&  0.91$\pm$0.12& 5.83         & 0.64$\pm$0.10&  1.32$\pm$0.17\\
HNCC$^{13}$C& 4-3&     &36007.309$\pm$0.010&  1.66$\pm$0.07& 5.83         & 1.11$\pm$0.05&  1.40$\pm$0.07\\ %6.6E9 R=87.9 %blended with C6H5CN
            & 5-4&     &45009.056$\pm$0.010&  1.10$\pm$0.11& 5.83         & 0.74$\pm$0.08&  1.40$\pm$0.14\\
DNCCC       & 4-3&     &35204.594$\pm$0.010&  1.60$\pm$0.07& 5.83         & 0.70$\pm$0.04&  2.26$\pm$0.08$^e$\\ %1.1E10 R=52.7
            & 5-4&     &44005.628$\pm$0.010&  1.23$\pm$0.10& 5.83         & 0.54$\pm$0.05&  2.13$\pm$0.14\\
\hline
\end{tabular}
%}\\
\tablefoot{
\tablefoottext{a}{Predicted or observed frequencies for the transitions of 
HCCNC and HNCCC and their isotopologs (see Sections \ref{spec_HCCNC} and \ref{spec_HNCCC}).
The observed frequencies have an  uncertainty of 10 kHz and have
been derived adopting a v$_{LSR}$ of 5.83 km\,s$^{-1}$ \citep{Cernicharo2020}. The term $Main$ refers to the $\Delta$$F$=+1 transitions
with $F_u=J, J+1, J-1$ which are at nearly the same frequency and unresolved in our data.}
\tablefoottext{b}{Integrated line intensity in mK\,km\,s$^{-1}$.} 
\tablefoottext{c}{Line width at half intensity using a Gaussian fit in the line profile (in km~s$^{-1}$).}
\tablefoottext{d}{Antenna temperature (in mK).}
\tablefoottext{e}{Data only come from the set of observations with frequency throw of 10 MHz.}
}\\
\end{table*}

\end{appendix}
\end{document}